\documentstyle[preprint,eqsecnum,aps]{revtex}
\begin{document}
\draft
\vskip 0.5cm
\preprint{McGill/94-35}
\title{A unified description for nuclear equation of state and
	fragmentation in heavy ion collisions}
\author{Jicai Pan \ and \ Subal Das Gupta }
\address{Department of Physics, McGill University \\
3600 University St., Montr\'{e}al, PQ, H3A 2T8 Canada}

\date{October 1994}
\maketitle
\begin{abstract}
 We propose a model that provides a unified description of nuclear
equation of state and fragmentations. The equation of
state is evaluated in Bragg-Williams as well as in
Bethe-Peierls approximations and compared with that in the
mean field theory with Skyrme interactions. The model shows a
liquid-gas type phase transition. The nuclear fragment
distributions are studied for different  densities at finite
temperatures. Power law behavior for fragments is observed at critical
point. The study of fragment distribution and the second moment
$S_2$ shows that the thermal critical point coincides with the percolation
point at the critical density.  High temperature behavior of the model
shows characteristics of chemical equilibrium.

\end{abstract}

\newpage
\section{Introduction}

The phenomenon of nuclear matter fragmenting into various
pieces can be studied in heavy ion collisions.
This has been an area of much activity, both in
experiments and in theory.  Curtin, Toki and Scott \cite{curtin}
pointed out that at some incident energies excited matter
that is formed in heavy ion collisions
will pass through a liquid-gas phase transition stage
and if fragments are formed at this stage, it may show
characteristics of this transition.  A study
of nuclear matter with a Skyrme interaction
was made in \cite{jaquaman} where it was shown that in mean field
theory there is a phase transition
as in Van der Waals gas. In nuclear physics, phase-transition,
if it indeed happens, is a transient phenomenon and it is
not clear what the mean field theory predicts for observables
that can actually be measured.
In experiments the most readily accessible observable
is the cross section of nuclear fragments,
or yield $Y(A)$ $vs.$ $A$ where $A$ is the mass number
or charge of a fragment.
Of course mean field theories can not make a prediction
about fragment distribution thus it falls
short of directly providing results with which
one could confront the data.

Bauer \cite{bauer1} and Campi \cite{campi1} used the
percolation model \cite{stauffer} to calculate fragment
distribution \cite{bauer2,campi2}.  There are two varieties of
percolation models: the bond and the site percolation.
In a bond percolation model each site of the lattice
is occupied by a nucleon. That is, the number of
nucleons equals the number of lattice sites $N$.
The bonds between nearest neighbors are broken with a
probability of $1-p$ where $p$ is the probability that the
bond is unbroken.  Nucleons which are connected through
unbroken bonds form a cluster.  In a site percolation model,
each site is occupied randomly with probability $p\leq 1$.
The number of nucleons,  $n$, is usually less or equal to
the number of the lattice sites $N$.
Nucleons occupying nearest neighbor sites
are considered to be the part of one cluster.
In both bond and site percolation
models there is a critical value $p_c$ of $p$ at
which an infinite cluster starts to emerge.
For very large lattice $p_c$ is independent
of $N$.  The probability that a given
site belongs to this infinite cluster is zero for $p<p_c$, and
grows from 0 to 1 for $p\geq p_c$.
In general, the cluster distributions in
percolation models  are very similar to the mass distribution
observed in heavy ion collisions \cite{bauer2,campi2}.

The percolation model is quite different from mean field
theories that we first alluded to.  There is just one parameter
in a percolation model. In a bond percolation model the
number of sites equals the number of nucleons.
One might regard that the probability $p$ is a function of
temperature.  In that case there is no parameter corresponding
to the volume or the pressure of the system. If instead one
takes the site percolation model, the number of lattice sites
$N$ is usually larger than $n$. One may now associate the
parameter $p$ with the volume.  There is, then, no reference to
temperature or pressure.  Thus the simple
percolation model can not describe the thermodynamic
aspect of nuclear fragmentation.

Our present work started with the desire to have a model that
can both describe the equation of state and the fragmentation
of finite nuclear systems. The quintessence of this model
is the lattice gas model \cite{huang}.
The model leads to a $(P,V,T)$ diagram, and it is also linked
with the percolation model in an obvious way.  An interesting
feature is that it not only leads to a liquid-gas type of phase
transition, but also encompasses the percolation transition.

Much work has been published on the subject of
liquid-gas phase transition in nuclei.
See Refs. \cite{grossa,siemens,lopez}
for description of some early works and \cite{csernai} for review.
There are microscopic models which are proposed to calculate
multifragmentation \cite{bauer3,gallego,chomaz,gross,ayik}.
Other phenomenological models can
be found in \cite{chase}.  The pioneering work by the Purdue group
\cite{finn} was a big impetus for this subject.

\section{The Model}

We consider the participant zone in collisions of two heavy ions
and conjecture that because of nucleon-nucleon collisions the
system reaches thermal equilibrium.  We assume that the system
is adequately described by classical statistical mechanics.  In
that case the canonical partition function of an $n$-particle
system can be written in a separable form:
$Z_{can}\propto Z_p(can)Z_r(can)$ where $Z_p(can)$ is given by
\begin{eqnarray}
Z_p(can)=\int \exp[-\beta \sum_1^np_i^2/2m]
	d^3p_1......d^3p_n \label{z1}
\end{eqnarray}
where $\beta $ is the inverse temperature.  The other part of the
partition function is
\begin{eqnarray}
Z_r(can)\propto \int \exp[-\beta\sum_{i<j}v(r_{ij})]
	d^3r_1....d^3r_n \label{z2}
\end{eqnarray}
Here $v(r_{ij})$ is the potential between particles $i$ and $j$.
We approximate the configuration part of the partition function
by the partition function of the lattice gas model.  In this model
(as in a site percolation model) each lattice site can be occupied by
at most one nucleon. The number of lattice sites, $N$, gives a measure
of the volume of the system and is usually larger than $n$.
When the nucleus is in the ground state we have $n=N$. Thus our
model is limited to normal nuclear volume or higher.  Becuase cluster
formation presumably takes place at a volume much larger than
normal volume, this restriction may not be a debilitating factor.
In contrast with the
percolation model, the lattice gas model includes interactions.
If two nearest neighboring sites are filled by
nucleons they will interact and the energy of interaction is
denoted by $-\epsilon $.  The nearest
neighbor interaction simulates short range nature of nucleon-nucleon
interaction.

We can now write down the canonical partition function for our model
for a fixed $n$ and $N$. The $Z_p(can)$ is simple and does not depend
on the volume, and we just need to calculate the $Z_r(can)$.
Let $N_{nn}$ be the number of $nn$ bonds in a specific lattice
configuration, the energy carried by these bonds is then $-\epsilon N_{nn}$.
Then $Z_r(can)$ is given by
\begin{eqnarray}
Z_r(can)=\sum_{N_{nn}}g(N,n,N_{nn})
	e^{\beta \epsilon N_{nn}} \label{z3}
\end{eqnarray}
where $g(N,n,N_{nn})$ is a degeneracy factor satisfying
\begin{eqnarray}
\sum_{N_{nn}}g(N,n,N_{nn})=\frac{N!}{(N-n)!n!} \label{g}
\end{eqnarray}
An exact evaluation of $Z_r(can)$ given in (\ref{z3}) is
difficult. Thus we will use approximate means.

\section{Equation of state}
\subsection{Bragg-Williams approximation}

The Bragg-Williams approximation is an easy and quick calculation
but is not expected to be accurate. The results are, however, transparent
and analytically demonstrate Van der Waals type behavior.
In this approximation the number of $nn$ bonds
$N_{nn}$ is taken to be given and fixed, when $N$ and $n$ are specified.
If one site is definitely occupied, the number of its $\gamma$
neighbors that are occupied is on the average $\gamma n/N$.
In our three dimensional simple cube lattice $\gamma$ is 6 except at the
boundary. Since there are $n$ nucleons in the system, the number of $nn$
bonds is then $\gamma n^2/2N$ where we
have ensured that each bond is only counted once and
assumed that both $n$ and $N$ are large so that boundary effects
can be neglected.
In Bragg-Williams approximation the canonical partition
function is then
\begin{eqnarray}
Z_r(can)=\frac{N!}{(N-n)!n!}
e^{\frac{1}{2}\beta\epsilon\gamma\frac{n^2}{N}}. \label{bwz}
\end{eqnarray}
The equation of state can be calculated by utilizing
$P=kT(\partial \ln Z(can)/\partial V)_T=kT(\partial
\ln Z_r(can)/\partial V)_T$ since $Z_p(can)$ does not
have any $V$ dependence.
Here $P$ is the pressure and $V$ is
the volume given by $V=a^3N$. A representative value of
$a^3$ would be $a^3=1/\rho_0=6.25$ fm$^3$ where $\rho_0$
is the normal nuclear density. The normal nuclear volume is
$V_0=a^3n$. Using Stirling's approximation
for $N!$, $n!$ and $(N-n)!$ one can show
\begin{eqnarray}
P=\frac{kT}{a^3}\ln\frac{N}{N-n}-
  \frac{1}{2a^3}\epsilon\gamma (\frac{n}{N})^2.
\end{eqnarray}
Using $n/N=V_0/V=\rho/\rho_0$ we finally get
\begin{eqnarray}
P=kT\rho_0\ln\frac{V}{V-V_0}-
 \frac{1}{2}\epsilon\rho_0\gamma(\frac{V_0}{V})^2. \label{bwp}
\end{eqnarray}
This equation of state has the same qualitative behavior as the
Van der Waals gas.  For one mole of gas, the Van der Waals
equation of state is
\begin{eqnarray}
P=\frac{N_AkT}{V-b}-\frac{a}{V^2} \label{van}
\end{eqnarray}
The lattice gas pressure goes to infinity as $V$ approaches
$V_0$.  The Van der Waals gas pressure goes to infinity as $V$
is squeezed to the value $b$.  For large $V$ both the equations
of state approach the perfect gas limit.
The critical point can be readily
determined analytically from (\ref{bwp}). By setting
$\partial P/\partial \rho=\partial^2P/\partial \rho^2=0$ at
the critical point we obtain
$\rho_C=0.5\rho_0$ and $kT_C=\gamma\epsilon/4$, respectively.  It is
also straightforward to show that in the Bragg-Williams
approximation $P_CV_C/RT_C=2\ln2-1=0.386$ for the lattice gas.
The corresponding number for the Van der Waals gas is 0.375.

The Bragg-Williams approximation is considered to be crude and one may
wonder if the lattice gas model would indeed lead to a liquid-gas type
phase transition when a better mean-field calculation is done.  In the
next subsection we try an improved approximation.

\subsection{Bethe-Peierls approximation}

We now try to do a better mean field calculation and use
what is called the Bethe-Peierls approximation in the Ising model.
For the Ising model, the order parameter can be computed without
having to calculate the partition function but for the
equation of state that we wish to calculate we will need to obtain the
partition function.
Here we consider a grand canonical ensemble.  To explain the
methodology we refer to Fig. 1 where, for simplicity, a
two dimensional square lattice is shown.
We break up the lattices into $N/(\gamma+1)$ blocks, each of which
contains $\gamma+1$ sites.  The
interactions within each block are treated exactly while
the interactions between different blocks are taken
into account approximately. The local correlations are taken into
account in this approximation, and it is an improvement over
Bragg-Williams approximation. The grand partition function can
be written as the product of the grand partition functions of the
$N/(\gamma+1)$ blocks:
\begin{eqnarray}
Z_{gr}=z_{gr}(\mbox{block 1})z_{gr}(\mbox{block 2})\cdots
 z_{gr}(\mbox{block }\frac{N}{\gamma+1}). \label{bp1}
\end{eqnarray}
The grand partition function of the block labeled by $1,2,3,\gamma$ and 5
can be written as
\begin{eqnarray}
z_{gr}=\sum_{k=0}^{\gamma +1}\sum_{\alpha} e^{\lambda'k}
	e^{-\beta E_{\alpha}(k)} \label{bp2}
\end{eqnarray}
Here $\lambda '$ plays the role of chemical potential.
The energy $E_{\alpha}(k)$ consists of two parts; the kinetic energy
($\sum_{i=1}^k p^2_i/2m$) and the potential energy for nearest neighbor
interaction. Integration over the kinetic
energy part can be immediately done to give a factor of
$[(2\pi m/\beta )^{3/2}]^k \equiv e^{qk}$.
We now define $\lambda =\lambda '+q$ and also divide the
the right hand side of the above equation into two parts:
\begin{eqnarray}
z_{gr}=\sum_{k=0}^{\gamma}{\gamma\choose k}
 e^{\lambda k+\beta k \overline{\epsilon }}+
 e^{\lambda}\sum_{k=0}^{\gamma}{\gamma\choose k}e^{\lambda k+
\beta k(\epsilon+\overline{\epsilon})}. \label{bp3}
\end{eqnarray}
The first part is proportional to
the probability that site 5 is unoccupied and its neighboring
sites $1\cdots\gamma$ are $k$-fold occupied where $k$ goes from
0 to $\gamma$.  The second part is proportional to the probability
that site 5 is occupied where an extra amount of energy $-k\epsilon$
is included when $k$ of the nearest neighbors are also occupied.
The extra factor $\exp[\beta k\overline{\epsilon }]$
takes into account the interaction between different blocks.
If we left this factor out the nucleon at site 1 (Fig. 1) would only
interact with the one at site 5.

Because of self-consistency condition the average occupation
probability at every site must be the same.
The average occupation at site 5 is
\begin{eqnarray}
P(5)=\frac{e^{\lambda}(1+e^{\lambda+
 \beta \epsilon +\beta \overline{\epsilon }})^
{\gamma}}{z_{gr}}=\frac{n}{N} \label{bp4}
\end{eqnarray}
where
\begin{eqnarray}
z_{gr}=(1+e^{\lambda+\beta \overline{\epsilon }})^{\gamma}+
 e^{\lambda}(1+e^{\lambda+
\beta \epsilon +\beta \overline{\epsilon }})^{\gamma} \label{bp5}
\end{eqnarray}
is obtained from (\ref{bp3}) by summing over $k$.
The average number of particles in all the sites neighboring
site 5 is
\begin{eqnarray}
\sum_{i=1}^{\gamma}P(i)
=\frac{\sum{\gamma\choose k}k
e^{\lambda k+\beta k\overline{\epsilon}}+
e^{\lambda}\sum{\gamma \choose k}k e^{\lambda k+
\beta k\epsilon+\beta k\overline{\epsilon}}}{z_{gr}}. \label{bp6}
\end{eqnarray}
The self-consistency condition then implies
\begin{eqnarray}
e^{\lambda}(1+e^{\lambda+\beta\epsilon+
\beta\overline{\epsilon}})^{\gamma}=
e^{\lambda+\beta\overline{\epsilon}}
(1+e^{\lambda+\beta\overline{\epsilon}})^{\gamma-1}
+e^{2\lambda+\beta(\epsilon+\overline{\epsilon})}
(1+e^{\lambda+\beta(\epsilon+\overline{\epsilon})})^{\gamma-1}. \label{bp7}
\end{eqnarray}
The two unknowns $e^{\lambda}$ and $\overline{\epsilon}$ can
be solved from (\ref{bp4}), (\ref{bp5}) and (\ref{bp7}).

Divide both sides of (\ref{bp7}) by
$e^{\lambda}(1+e^{\lambda+
\beta(\epsilon+\overline{\epsilon})})^{\gamma-1}$,  we obtain
\begin{eqnarray}
1=e^{\beta\overline{\epsilon}}
\left[\frac{1+e^{\lambda+\beta\overline{\epsilon}}}
{1+e^{\lambda+\beta\epsilon+\beta\overline{\epsilon}}}
\right]^{\gamma-1}, \label{bp8}
\end{eqnarray}
and (\ref{bp4}) can be rewritten as
\begin{eqnarray}
\frac{N}{n}=1+\frac{(1+\lambda
e^{\beta\overline{\epsilon}})^{\gamma}}{e^{\lambda}(1+
e^{\lambda+\beta\epsilon+\beta\overline{\epsilon}})^{\gamma}}. \label{bp9}
\end{eqnarray}
Use (\ref{bp8}), (\ref{bp9}) becomes
\begin{eqnarray}
\frac{N}{n}=1+\frac{1}
{e^{\lambda+\beta\overline{\epsilon}\gamma/(\gamma-1)}}.\label{bp10}
\end{eqnarray}
We can now write $e^{\lambda}$ in terms of $\overline{\epsilon}$
\begin{eqnarray}
e^{\lambda}=\frac{n}{N-n}
 e^{-\beta\overline{\epsilon}\gamma/(\gamma-1)}. \label{bp11}
\end{eqnarray}
Combining (\ref{bp4}), (\ref{bp5}) and (\ref{bp11}) we obtain
\begin{eqnarray}
x=\frac{1}{2}\left[\frac{N-2n}{N-n}+\sqrt{(\frac{N-2n}{N-n})^2+
4(\frac{n}{N-n})e^{\beta\epsilon}} \right] \label{bp12}
\end{eqnarray}
where
\begin{eqnarray}
x=e^{\beta\overline{\epsilon}/(\gamma-1)}. \label{bp13}
\end{eqnarray}
The values of $\overline{\epsilon}$ and $e^{\lambda}$ can now be found
from (\ref{bp11}) -- (\ref{bp13}). We note here that the results depend
on the ratios of $n/N$ and $\epsilon/kT$ only.

Let us now go back to the partition function for the lattice
given in (\ref{bp1}) where it is written as a product of the partition
functions of the $N/(\gamma+1)$ blocks. If we simply use the
partition function for each block given in (\ref{bp5})
we count twice the binding energies between neighboring sites in
different blocks.   For example, the binding
energy between 1 and 6 (Fig. 1) is included in
$z_{gr}(\mbox{block 1})$ through
$\overline{\epsilon}$, and it is included again in
$z_{gr}(\mbox{block 2})$.
We note that on the average there are $n/N$ particles
at each site, and each block has $\gamma$ peripheral sites.
Thus, when we evaluate the partition function for
the lattice, the partition function for each block given in
(\ref{bp5}) should be corrected with following factor:
\begin{eqnarray}
\mbox{correction}=
  e^{-\frac{1}{2}\beta\overline{\epsilon}\gamma n/N}.\label{bp14}
\end{eqnarray}
We can now use $PV/kT=\ln Z_{gr}$, $V=N/\rho_0$ and
$\ln Z_{gr}=N/(\gamma+1)\ln z_{gr}$ to obtain
\begin{eqnarray}
P=\rho_0kT\times \frac{1}{\gamma+1}\ln z_{gr}. \label{bp15}
\end{eqnarray}
Here $z_{gr}$ is now understood as the product of a block
partition function given in (\ref{bp5}) and the correction
factor given in (\ref{bp14}).

The equations of state calculated in Bragg-Williams
approximation and Bethe-Peierls are compared in Fig. 2
in $P-V$ diagrams at different temperature.
It is seen that in high temperature limit both give the
same results, and they begin to differ at low temperatures.

\subsection{Mean field theory with Skyrme interaction}

We take a Skyrme interaction with potential energy density
given by \cite{jaquaman}
\begin{eqnarray}
V(\rho)=\frac{A}{2}\rho_0(\frac{\rho}{\rho_0})^2+
 \frac{B}{\sigma+1}\rho_0(\frac{\rho}{\rho_0})^{\sigma+1}  \label{sk1}
\end{eqnarray}
where $A=-356$ MeV, $B=303$ MeV and $\sigma$=7/6.
This interaction produces a saturation density of
$0.16$ fm$^3$ and binding energy of 16 MeV per particle.

The pressure consists of two parts: one originating from the
interaction and the other from the kinetic energy.
The pressure produced by the
interaction is given by
\begin{eqnarray}
\left[\frac{A}{2}\frac{\rho}{\rho_0}+
\frac{\sigma B}{\sigma+1}
(\frac{\rho}{\rho_0})^{\sigma}\right]\rho .  \label{sk2}
\end{eqnarray}
The kinetic pressure is calculated
numerically from a Fermi gas model at finite temperature.

 Similar to Fig. 2 one can draw $P$ against $V/V_0$ ($=\rho_0/\rho$)
at various
$T$. A comparison of the equation of state in
the mean field theory with Skyrme interaction and in the
lattice gas is shown in Fig. 3. In both cases we see
characteristics of liquid-gas phase transition. In the calculation
we used $\epsilon$=9 MeV and $\rho_0=0.16$ fm$^{-3}$.

\section{Fragment distributions}

  Nuclear fragmentations are described by the formation of clusters
in our model. To generate fragements we need to simulate the lattice
configuration and momenta of particles, and to determine if neighboring
particles belong to the same cluster. The lattice configuration is
generated according to the (\ref{z2}), while the momenta are generated
from Maxwell-Boltzmann factor given in (\ref{z1}). These
two processes are independent of each other since the partition function
can be factorized. To generate the configuration we start from an empty
lattice and put the first nucleon at random.  Once this has been put in
the $\gamma $ boxes that are immediate neighbours are assigned
a probability $\propto \exp[\beta \epsilon]$ whereas all other
boxes have probability proportional to unity.  The next nucleon
is then put in according to this probability distribution.  If
at an intermediate state there are $m$ empty boxes we assign to
each of these boxes a probability proportional to
$\exp[q\beta \epsilon]$ where $q$ is the number of nearest
neighbors that are already filled up.  The next filling is then
done according to this distribution.  The difference from the
computer simulation that would be done in a site percolation model
is the Boltzmann factor.  Having done the configuration space sampling
we then assign each nucleon a momentum according to Maxwell-Boltzmann
distribution.

 In a site percolation model two neighboring particles
always form a cluster. The formation of a cluster in our model
depends on the interactions of neighboring particles and their
relative kinetic energy. We adopt the following physical criterion for
determining the formation of a cluster.  Two neighboring
nucleons belong to the same cluster if the following condition
is satisfied
\begin{eqnarray}
{\bf p}_r^2/2\mu-\epsilon<0 \label{a1}
\end{eqnarray}
Here ${\bf p}_r$ is the relative momentum between the two nucleons
and $\mu $ is the reduced mass.
Except at very low temperatures, the frequency with which two
nucleons appear at neighboring sites depends mostly on density.
The probability of ${\bf p}_r^2/2\mu$ exceeding the value $\epsilon $
increases with temperature since the momenta of each nucleon is
obtained from Monte-Carlo sampling of Maxwell-Boltzmann
distribution.  Hence the probability that
two nucleons are bonded decreases with increasing temperature
and the system becomes less compact at higher temperature as
it should.  A different parametrisation used in \cite{li1} leads to
similar effects.

We note that when each particle obeys the Maxwell-Boltzmann
distribution, the distribution of relative momentum between two
particles is also Maxwell-Boltzmann , i.e.,
$P({\bf p}_r)=1/(2\pi \mu kT)^{3/2} \ \exp[-{\bf p}_r^2/2\mu kT]$.
We can then write down a formula for bonding probability which is
temperature-dependent:
\begin{eqnarray}
p=1-\frac{4\pi }{(2\pi \mu kT)^{3/2}}\int_{\sqrt{2\mu\epsilon}}^{\infty}
e^{-{\bf p}_r^2/2\mu kT}p_r^2dp_r \label{a2}
\end{eqnarray}

Coniglio and Klein \cite{coniglio} used a different parametrisation
for the bonding prbability. They used
\begin{eqnarray}
p=1-\exp[-\beta \epsilon/2] . \label{a3}
\end{eqnarray}
This was mathematically devised so that
the thermal critical point would also be a percolation point, a feature,
as we shall see, is also present in our parametrisation.
A comparison of the two formulae is presented in Fig. 4.
An example of fragment distribution obtained in our simulation is
shown in Fig. 5. One could decipher from this figure that at
$T=0.5T_C$ the system percolates; at $T=1.5T_C$ and $T=2.0T_C$ there is
no percolation and that percolation sets in around $T=T_C$.

\section{Thermal critical point and percolation point}

Let $p_c$ be the value of $p$ at the percolation point,i.e., the point
when an infinite cluster just appears.
The cluster size distribution for a general $p$ can be parametrized as
\cite{stauffer};
\begin{eqnarray}
Y(A,p)=A^{-\tau}f\left[(p-p_c)A^\sigma\right] . \label{general}
\end{eqnarray}
The scaling function $f(x)$ can be determined by computer
simulations or experiments. In many cases, such as the Fisher model
\cite{fisher}, $f(x)$ is an exponential function. At the percolation
point, the fragment distribution obeys a simple
power law
\begin{eqnarray}
Y(A,p_c)\propto A^{-\tau} \label{tau}
\end{eqnarray}
At the percolation point the fluctuation
is the maximum. This means that if in the neighborhood of
the percolation point the yield $Y(A,p)$ is fitted by a power law,
the exponent $\tau$ will be a minimum at the percolation point.
Experimental data are often fitted to a power law and a minimum
in $\tau$ is searched to ascertain the percolation point \cite{li2}.

The second moment is defined as
$S_2=\sum A^2n(A)/n$ where $n(A)$ is the number of clusters
with $A$ nucleons and the sum excludes the largest cluster
(defined henceforth as $A_{max}$).  In the thermodynamic
limit $S_2$ diverges at the percolation point and obeys a
power law distribution
\begin{eqnarray}
S_2\propto |p-p_c|^{-\gamma} . \label{gamma}
\end{eqnarray}

  As mentioned, the order parameter given by
$\lim_{n\rightarrow\infty} A_{max}/n$ is zero for
$p<p_c$ in and increases with $p$ for $p\geq p_c$.
Above the percolation point the order parameter is given by
\begin{eqnarray}
A_{max}/n \propto |p-p_c|^{\beta} . \label{beta}
\end{eqnarray}
There are two independent critical exponents in
most statistical models and percolation models. One can readily show
that following relations exist.
\begin{eqnarray}
\gamma={3-\tau\over \sigma} ;
\end{eqnarray}
\begin{eqnarray}
\beta={\tau-2 \over \sigma}
\end{eqnarray}
and
\begin{eqnarray}
\tau=2+{\beta\over \beta+\gamma} .
\end{eqnarray}

For a finite system the second moment $S_2$ goes through a
maximum (instead of infinity), and $\tau$ goes through a minimum at
the percolation point. Thus $\tau$ and $S_2$ can be used to identify
percolation point in a finite system.

In our model there is, first of all, a thermal critical point which is
obtained at $\rho_C=0.5\rho_0$ and $T_C=1.1275\epsilon$
\cite{stinchcombe}.
This more exact value of $T_C=1.1275\epsilon$ is smaller than the value of
$1.5\epsilon$ that we obtained from Bragg-Williams approximation.  But, in
addition, there is a continuous range of percolation points.  Provided the
density is higher than a minimum value ($\approx .3\rho_0$) a percolation
point will be reached at a certain temperature.  The higher the density,
the higher the temperature at which percolation sets in.  If we identify
the point at which a maximum in $S_2$ or a minimum in $\tau$ is achieved
as the percolation point, then we see from figs. 6 and 7 that
at higher density percolation point is reached at a higher temperature.  One
can now ask the question: if a minimum in $\tau$ is seen in experiments
as in ref [19], one could interpret that as an indication that the percolation
point is reached but has it got any relevance with the thermal critical
point ?  As figure 7 shows if the freeze-out density is $0.5\rho_0$ then
the percolation point is reached at the thermal critical point.  It also
follows that if the freeze-out density is close to $0.5\rho_0$ then a minimum
in $\tau$ is obtained when the temperature of the dissociating system is
close to the critical temperature.

 We therefore have this remarkable result which was not $a$ $priori$
imposed.  In our model the thermal critical point is also a percolation
point.  Numerically this result can be explained by noting that at the
critical temperature the value of $p$ in our model differs little from
Coniglio-Klein parametrisation.  At density $\rho/\rho_0$=1, the system
begins to percolate at $T=1.47T_C$ in our model and at $T=1.55T_C$ in
Coniglio-Klein model.

\section{Multiplicity as a variable}

The implicit thinking in much of what is presented above is that in nuclear
collisions matter is compressed, heated up and we can talk of a freeze-out
density and temperature at the time of dissociation. Both the temperature
and the density are, however, not directly measurable.
In the past values of $\frac{3}{4}\rho_0$ and lower have been
used for the freeze-out density. It is not known accurately.
Given that the most easily measurable quantities are the
multiplicity (number of charged particles),$S_2$ and $A_{max}$, we ask:
can we determine the  freeze-out density  from the multiplicity dependence of
$S_2$ and $A_{max}$? In Figs. 8 and 9 we plot the
$S_2$ and the $A_{max}$ as functions of the multiplicity at different
densities. In the calculations, at a given density we took a
sufficient large range
of temperatures such that we cover the full range of multiplicity. For a given
multiplicity both $S_2$ and $A_{max}$ can vary from one event to another.
What is plotted is the average for a given multiplicity. From
Figs. 8 and 9 we can easily see the changes of $S_2$ and $A_{max}$
when the density is changed from $0.3\rho_0$ to $0.5\rho_0$.
The differences are however rather small between density $0.5\rho_0$ to
$1.0\rho_0$. Remembering that there will always be uncertainties in
experimental data due to contamination from pre-equilibrium particles,
spectators etc., we conclude that it is difficult from these
observations alone to determine the freeze-out density accurately.
Some other variables might better differentiate between different freeze-out
densities.

\section{Beam Energy as a variable}

Fig. 6 shows that it may be possible that freeze-out density
can be determined from
the temperature dependence of $\tau$. The temperature is not a
direct observable in experiments although it has often been deduced
indirectly from other data, notably from slopes of inclusive cross-sections.
Here we will first try to deduce the temperature from a simple theory so that
the temperature is given once the beam energy is given.  We consider the
experimental setup of \cite{li1}. One has nearly central collisions of two
nearly equal ions.  The experiment is carried out at various beam energies
in the laboratory.  We take the number of paticles to be 85 (corresponding to
central $^{40}$Ar+$^{45}$Sc collisions).
In a purely classical model the ground state
has no kinetic energy at zero temperature so that the ground state energy
per nucleon is $-\epsilon N_{nn}^{max}/n$ where $N_{nn}^{max}$ is the
maximum number of $nn$ bonds possible for particle number $n=85$.  Since
$N_{nn}^{max}$ is determined by geometry we can use experimental binding
energy ($\approx 8.5$ MeV/$n$) to fix the value of $\epsilon$.  At temperature
$T$ the average energy per particle is $1.5kT-\epsilon \overline{N}_{nn}/n$
where
$\overline{N}_{nn}$, the average value of $N_{nn}$ is obtained from computer
simulations.  We can then write
\begin{eqnarray}
\frac{3}{2}kT+\epsilon(N_{nn}^{max}-\overline{N}_{nn})=e^* \label{temp}
\end{eqnarray}
For equal mass non-relativistic nuclear collisions we have $e^*=E_{beam}/4$
where $E_{beam}$ is the beam energy per nucleon in the laboratory.
There is an implicit
assumption here that all available energy is converted to thermal energy.
Thus the temperature is related to the beam energy.  We can now fix
the freeze-out density at different values, obtain an effective $\tau $
at each beam energy and obtain points as in experiments \cite{li1}.  This is
shown in Fig. 10. However the fit with data is not good for any of the
densities employed.

This type of mapping between beam energy and temperature is not
accurate.  One of the sources of errors is the collective flow which is
known to account for some fraction of the available energy.  Better
mapping could be expected where temperature is deduced from other
experimental data \cite{jacak,ogilvie}.  In this approach, the tail of
the proton cross-section is fitted by assuming that the proton has
a Maxwell-Boltzmann distribution in a frame which is moving in the laboratory.
We take this mapping from \cite{li1}.
When the mapping from this phenomenological approach is used, the fit
with the experimental data is quite good when the freeze-out density is
taken to be $0.39\rho_0$ (see Fig. 11).  What is also very pleasing is that
the predictions for different freeze-out densities are also sufficiently
different to be experimentally accessible.
The combined study of $S_2$ and $\tau$ as a function of beam energy
should be useful in determining the freeze-out density.

\section{High temperature characteristics}

Most of the attention in the present work has been focussed to temperatures
that are close to what is believed to be the critical temperature of
nuclear matter.  Indeed, in the past the percolation model has
mostly been used for
mild to moderate excitation energies.  At higher energies (i.e., Bevalac
energies) other models have been used with moderate success.  These models
use approximations that are valid in the
high temperature/low density limits.  Chemical equilibrium between
different species is assumed.  We will call these models by a generic name,
the thermodynamic model.  In this model rather simple expressions for
the average number of monomers
(single nucleons$\equiv \overline{n}_1$), dimers (clusters of two
nucleons$\equiv
\overline{n}_2$), three body clusters ($\overline{n}_3$) etc. can be obtained.
An early review can be found in \cite{dasgupta}.  We do not expect
to find exact correspondence between these calculations and the
present classical model at high temperature limits
since we do not have quantum degeneracies.
The clusters in our model consisting of attached cubical boxes in the lattice
have
degeneracies also. Even for moderate sized clusters these degenaracies
require considerable effort to enumerate analytically.
Nonetheless, many features seen
in experiments are common in both the models.  For example, in the
thermodynamic model, $\overline{n}_1$,
$\overline{n}_2$ and $\overline{n}_3$ vary with temperature but in a way that
$T^{3/2} \overline{n}_2/(\overline{n}_1)^2$ remains constant.  Here we have
neglected the binding energy of the dimer with respect to $kT$.
Constancy for this ratio is obtained in our present model also.  Similarly
$\overline{n}_1\overline{n}_3/(\overline{n}_2)^2$ is a constant in both the
models.  The so called coalescence relation
$\frac {d^3\overline{n}_2}{d^3p}(2{\bf p}) \propto
(\frac {d^3\overline{n}_1}{d^3p}({\bf p}))^2$ is obeyed in both the
models.  Thus rather reasonable features emerge when the model is
extrapolated to the high temperature side.

\vskip -0.5cm
\section{Summary}
\vskip -0.5cm

We have presented  a model that has links with both mean field
aspects and fragmentation of nuclei.  We studied the equation of state
under different approximations. The model shows a liquid-gas type
phase transition. We also studied the nuclear fragmantaion and discussed
the critical exponents near critical or percolation points. Some comparisons
with experimental data were made.

 The purpose of this paper was to present the essentials of this model.
The issues we addressed here are far from
exclusive. Many other features may be further explored. The present model,
we believe, is one step forward from the percolation model which was
proved to be helpful for the analysis of experimental data.

\vskip -0.5cm
\section*{Acknowledgments}
\vskip -0.5cm

We would like to thank H. Guo,  C. Gale, J. Kapusta
and M. Grant for useful
discussions. This work was supported in part by the Natural
Sciences  and Engineering Research Council of Canada and
by the FCAR fund of the Qu\'ebec Government.

\newpage
\section*{Figure captions}
\begin{description}
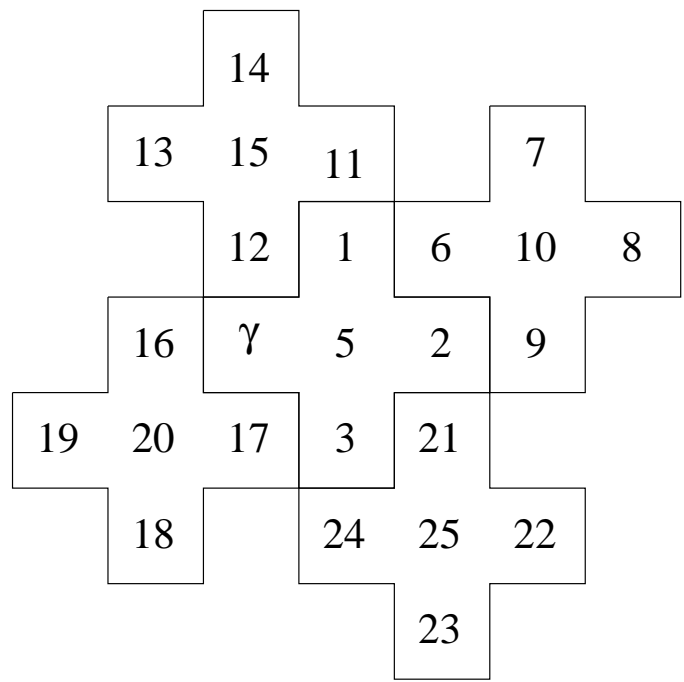
\item[Fig. 1] A square lattice is divided into blocks to illustrate
	the Bethe-Peierls approximations. See text for details.

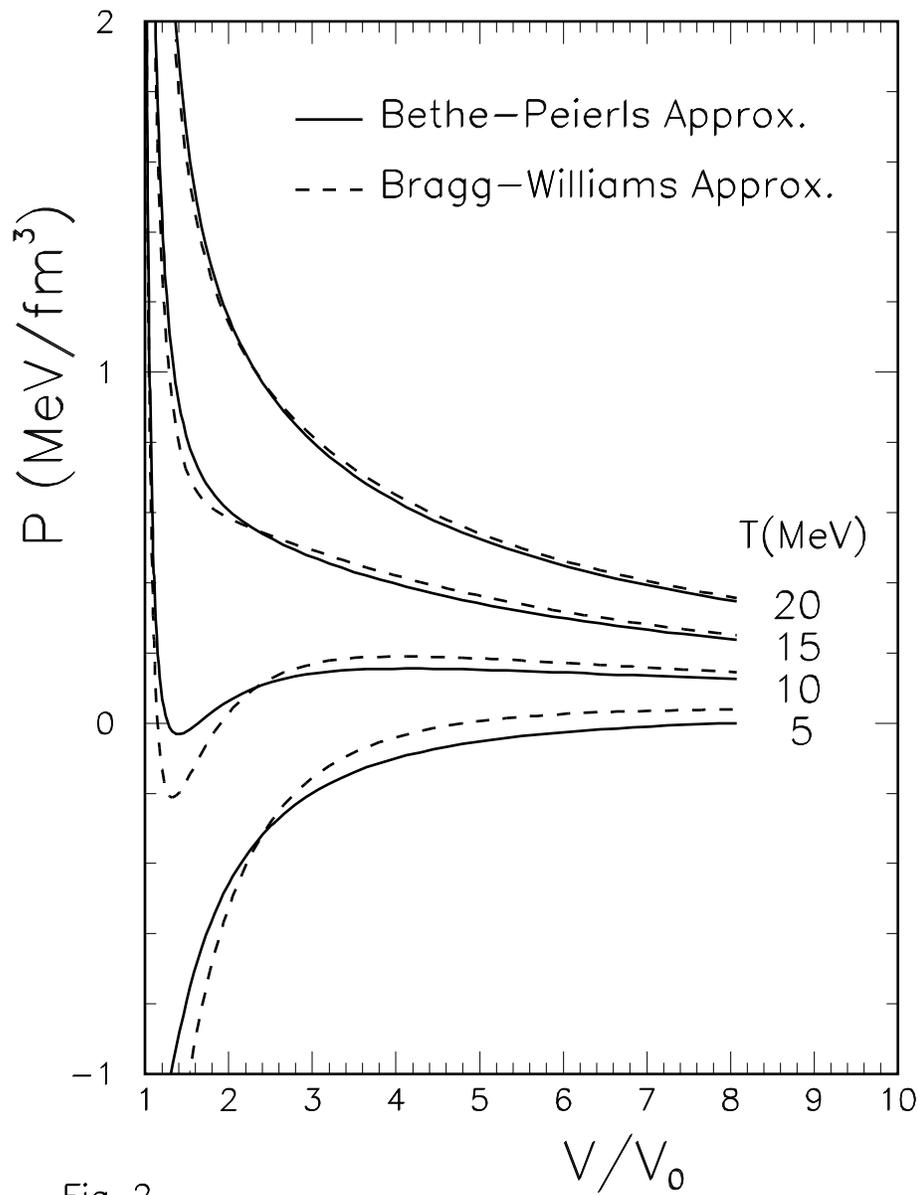
\item[Fig. 2] The $P-V$ diagrams in lattice gas model at
	temperatures $T=5$, 10, 15 and 20 (MeV).
	The solid curves are for the Bethe-Peierls
	approximation, and the dashed curves are for the
	Bragg-Williams approximation. In the calculation
	we used $\epsilon=9$ MeV.

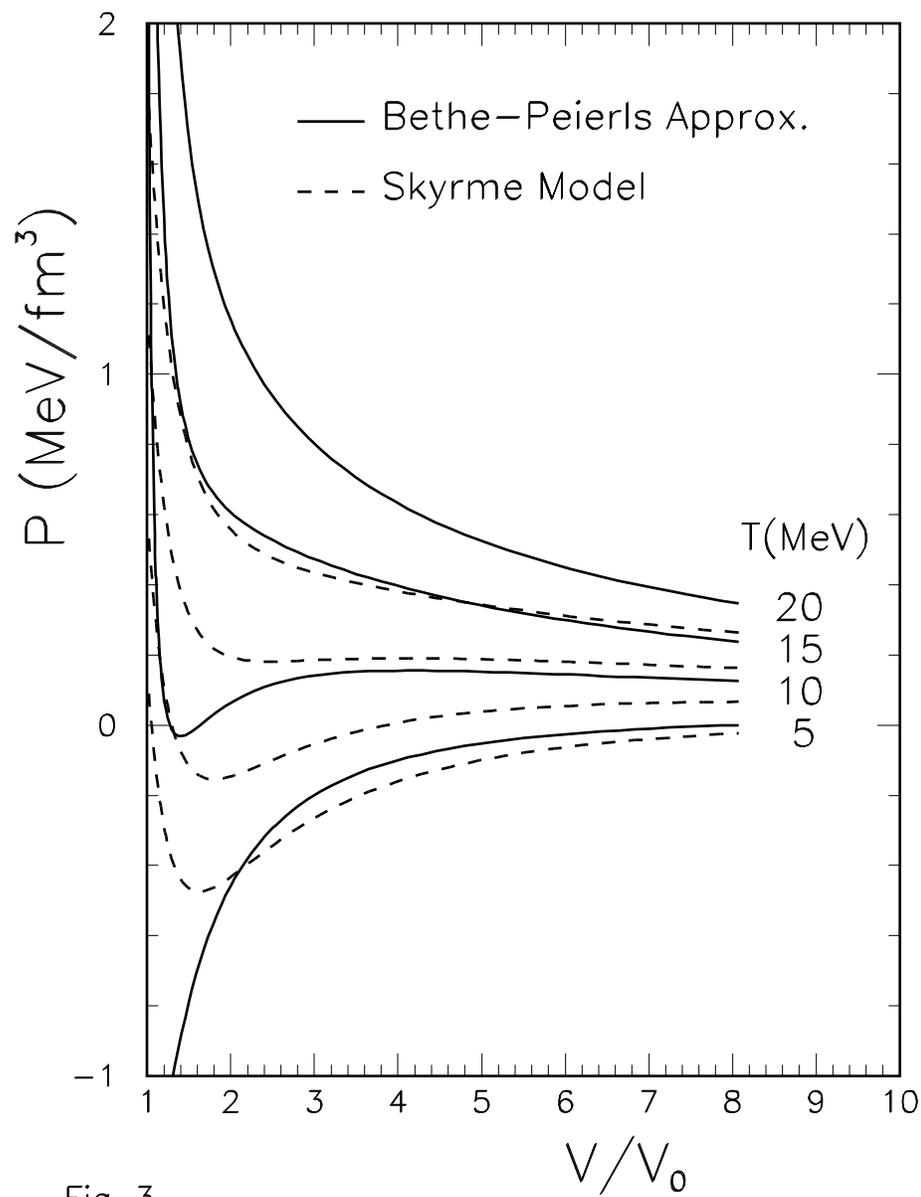
\item[Fig. 3] The same as Fig. 2, but the dashed curves are for the
	mean field theory with Skyrme interactions.

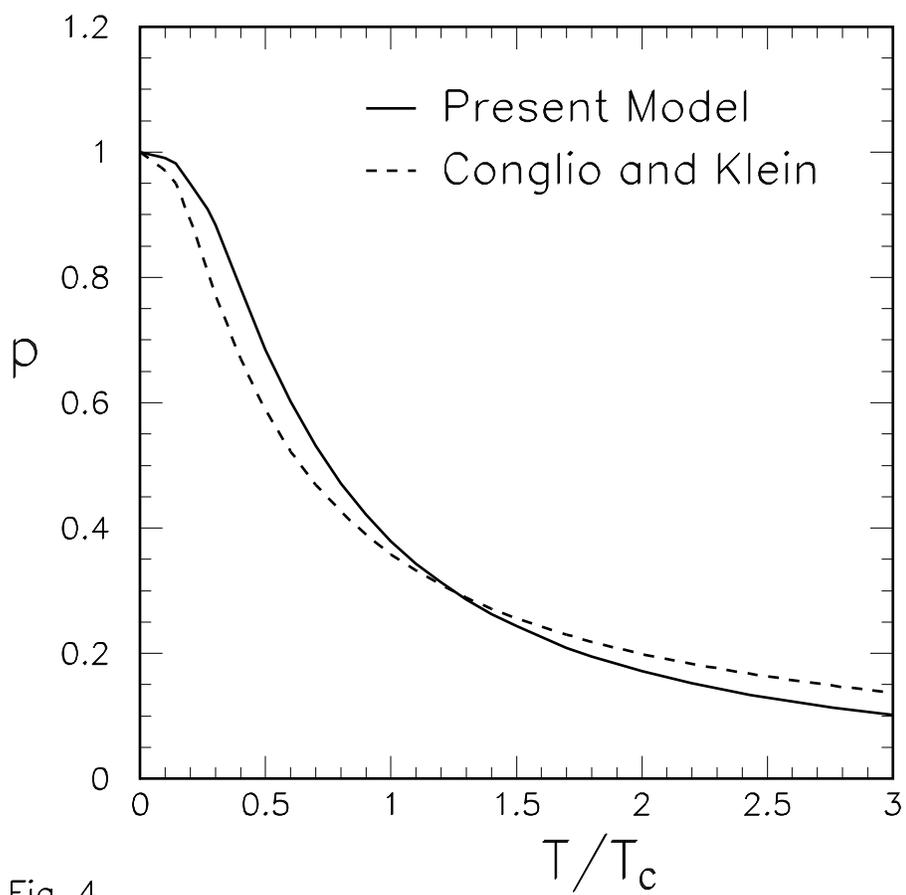
\item[Fig. 4] The bond probability $p_B$ is plotted as a function of
	temperature. The solid curve is obtained from our model given in
	eq. (\ref{a2}), and the dashed curve is the Coniglio and Klein model
	shown in eq. (\ref{a3}).

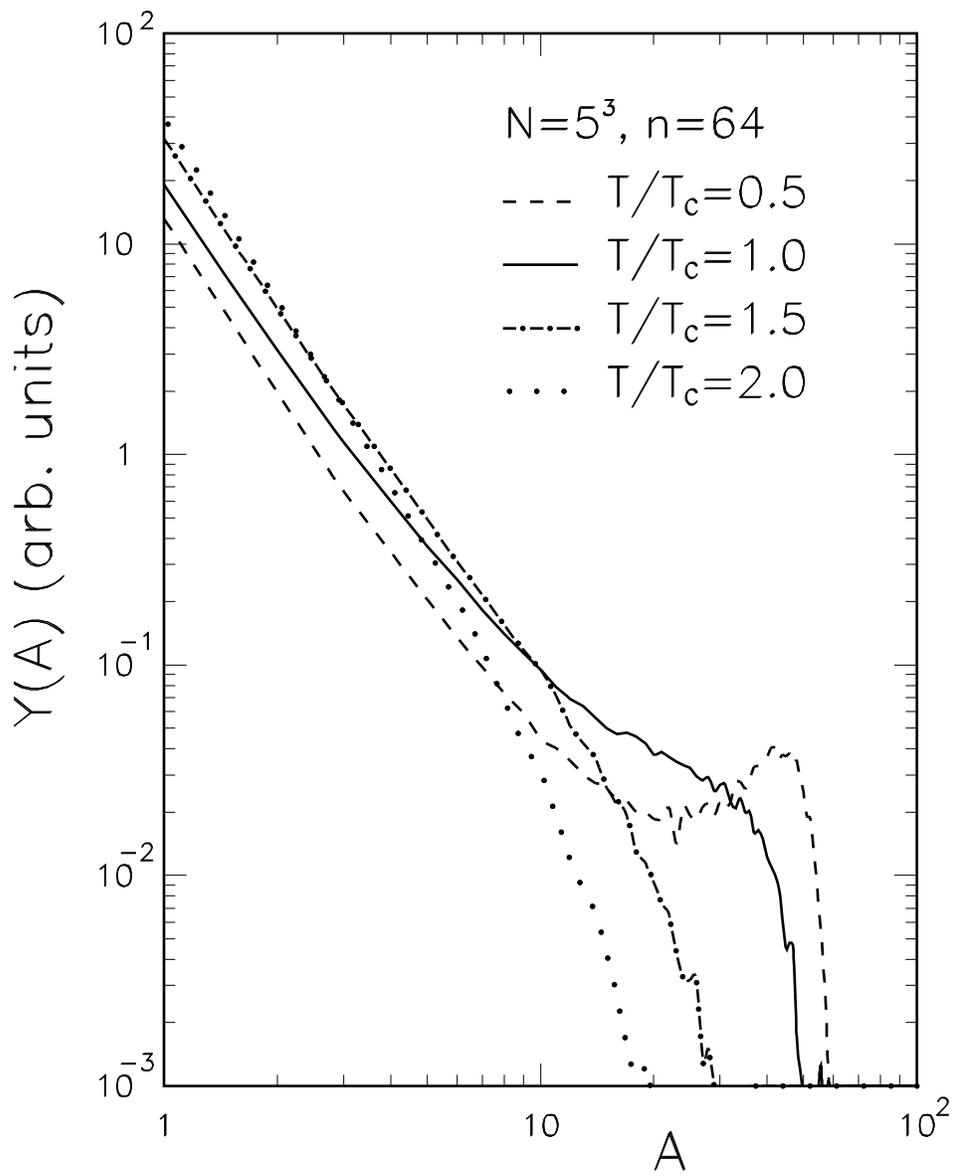
\item[Fig. 5] The mass yield distribution, $Y(A)$ $vs.$ $A$,
	for lattice $N=5^3$ and $n=64$, at temperatures
	$T/T_C=$0.5, 1.0, 1.5 and 2.0. Here $T_C=1.1275\epsilon$ is
	the thermal critical temperature. This value of $T_C$ taken from
	\cite{stinchcombe} is more accurate than the Bragg-Williams approximation.

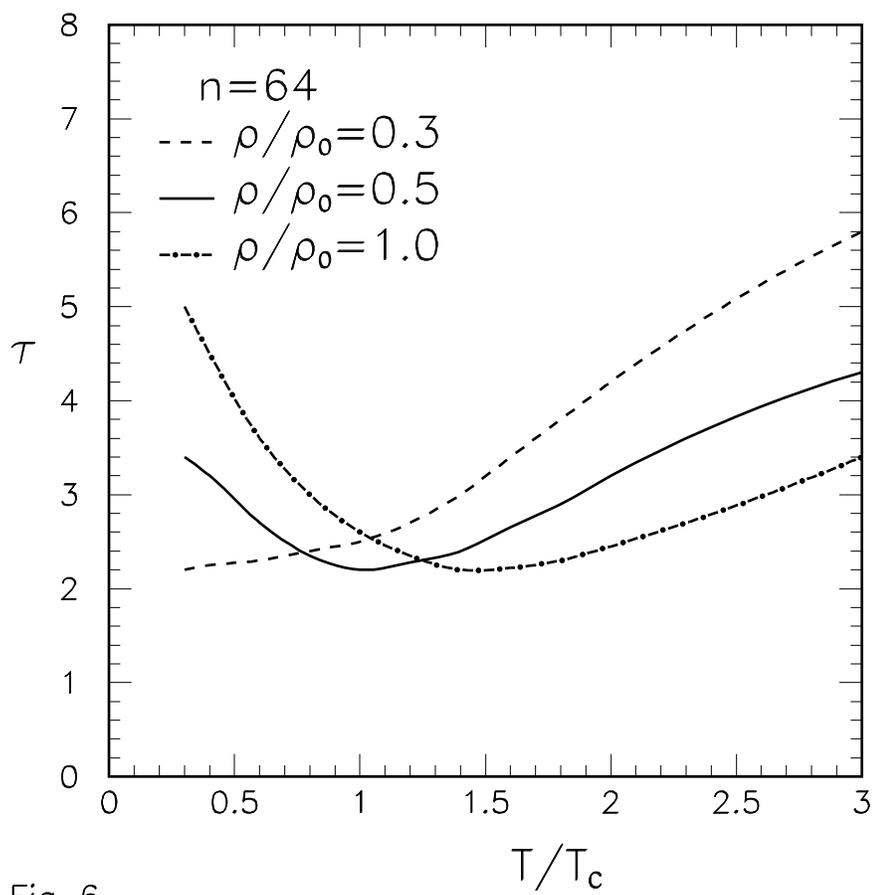
\item[Fig. 6] The value of $\tau$ is plotted as a function of temperature
	at different densities.

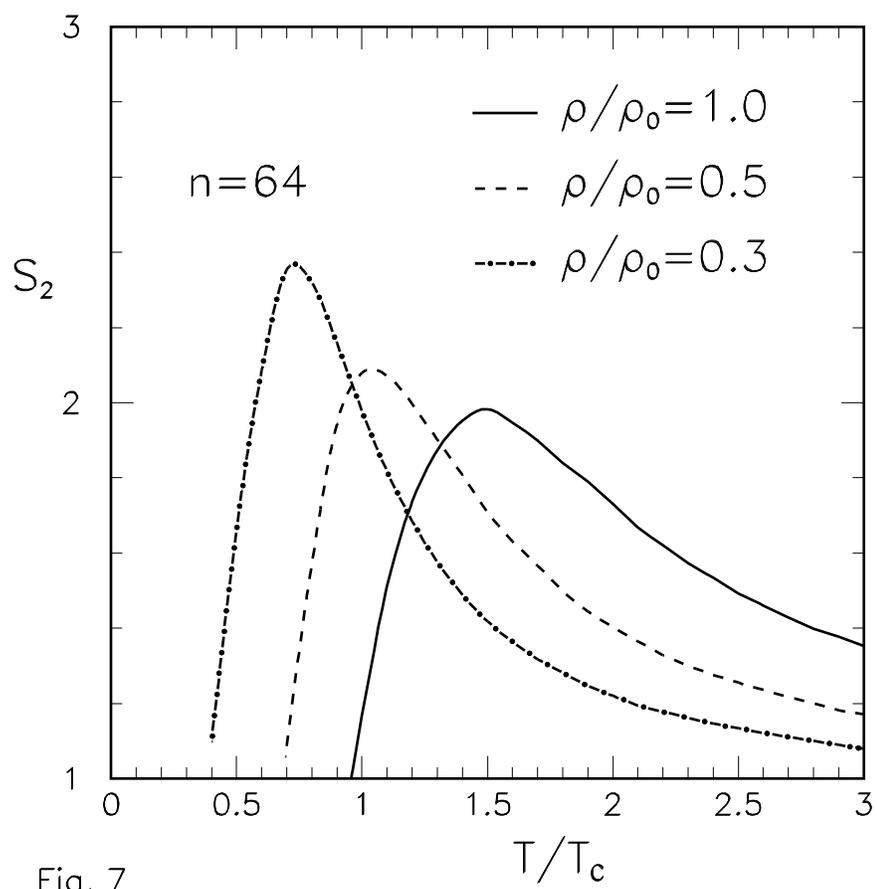
\item[Fig. 7] The value of $S_2$ is plotted as a function of temperature
	at different densities.

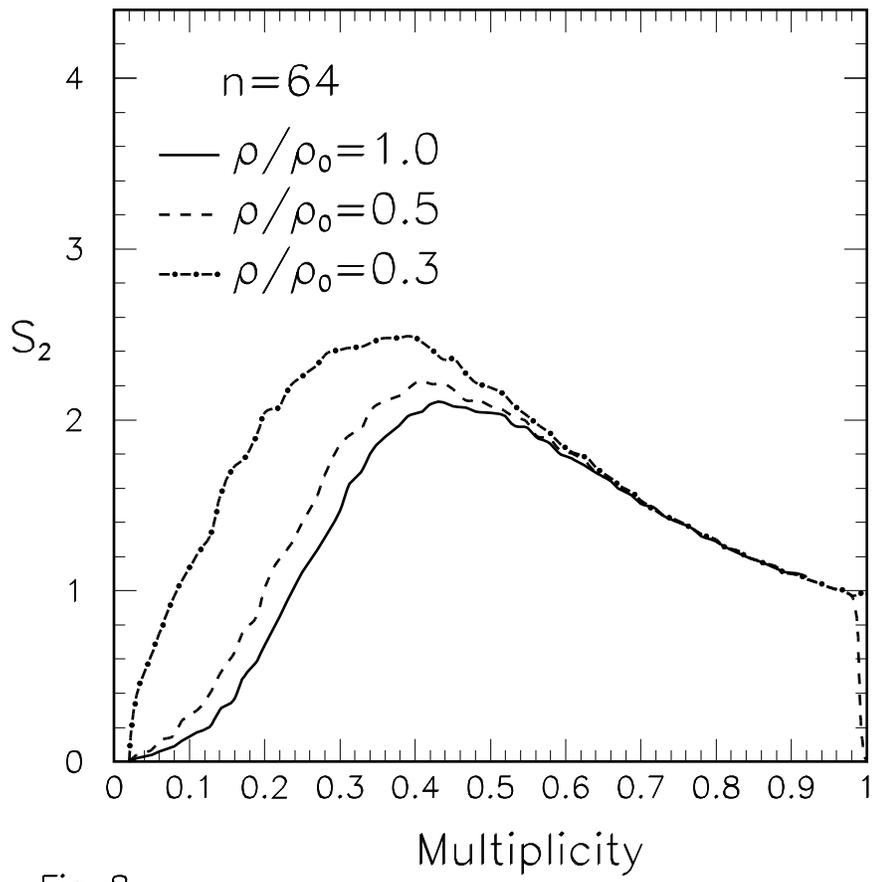
\item[Fig. 8] The value of $S_2$ is plotted as a function of multiplicity of
	fragments at different densities.

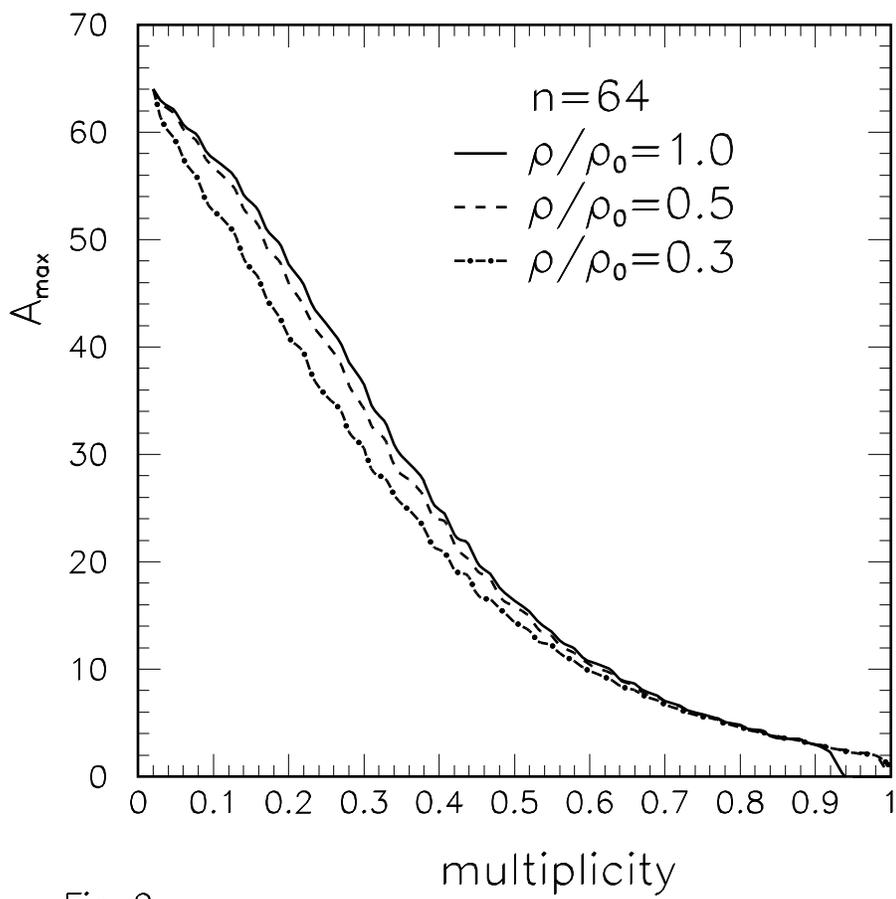
\item[Fig. 9] $A_{max}$ is plotted as a function of multiplicity of
	fragments at different densities.

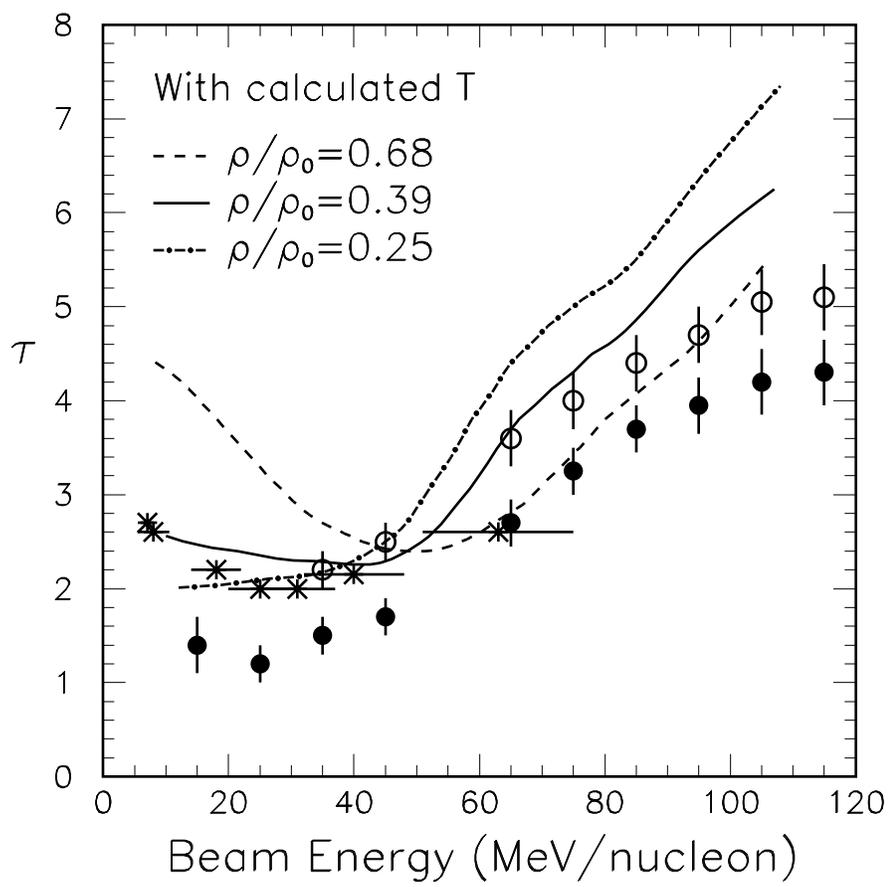
\item[Fig. 10] The theoretical exponent $\tau$ is compared with experimental
data
at different beam energies. The curves are obtained by using the
temperature calculated from eqs. (\ref{temp}).
The solid circles are the corrected data taken from \cite{li2},
the open circles are the uncorrected data taken from \cite{li1},
and the crosses are taken from \cite{ogilvie}.

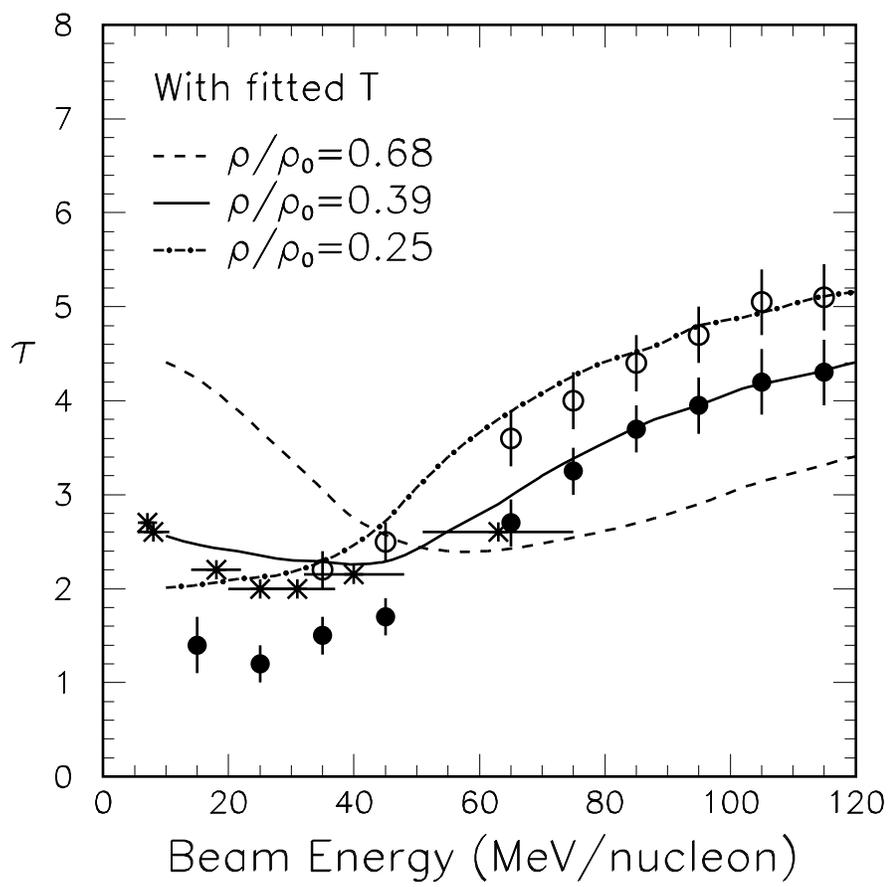
\item[Fig. 11] The same as Fig. 10, but the curves are obtained by using the
temperature fitted from experimental data.

\end{description}


\newpage
\begin{thebibliography} {99}

\bibitem{curtin} M. W. Curtin, H. Toki and D. K. Scott,
	Phys. Lett. {\bf B123}, 289 (1983).

\bibitem{jaquaman} H. Jaquaman, A. Z. Mekjian and L. Zamick,
	Phys. Rev. {\bf C27}, 2782 (1983).

\bibitem{bauer1} W. Bauer, D. R. Dean, U. Mosel and U. Post,
	Proc. seventh High Energy heavy ion study
	(GSI Darmstadt, 1984), GSI Report-85-10, p.701.

\bibitem{campi1} X. Campi and J. Debois, Proc. seventh
	High energy heavy ion study (GSI Darmstadt, 1984),
	GSI Report-85-10, p.707.

\bibitem{stauffer} D. Stauffer and A. Aharony, Introduction to
	percolation theory (Taylor and Francis, London, 1992).

\bibitem{bauer2} W. Bauer, Phys. Rev. {\bf C38}, 1297 (1988).

\bibitem{campi2} X. Campi, Phys. Lett. {\bf B208}, 351 (1988).

\bibitem{huang} K. Huang, Statistical Mechanics
	(John Wiley and Sons, Toronto,second edition, 1987).

\bibitem{grossa} D.H.E. Gross, L. Satpathy, T. Meng and M. Satpathy,
        Z. Physik C {\bf 309}, 42 (1982).

\bibitem{siemens} P. J. Siemens, Nature {\bf 305}, 410 (1983).

\bibitem{lopez} J. A. Lopez and P. J. Siemens,
	Nucl. Phys. {\bf A431}, 728 (1984).

\bibitem{csernai} L.P.Csernai and J.I. Kapusta,
	Phys. Rept. {\bf 131}, 223 (1986).

\bibitem{bauer3} W. Bauer, G.F.Bertsch and S. Das Gupta,
	Phys. Rev. Lett. {\bf 58}, 863 (1987).

\bibitem{gallego} J. Gallego, S. Das Gupta, C. Gale, S. J. Lee,
	C. Pruneau and S. Gilbert, Phys. Rev. {\bf C44}, 463 (1991).

\bibitem{chomaz} Ph. Chomaz, G. F. Burgio and J. Randrup,
	Phys. Lett. {\bf B254}, 340 (1991).

\bibitem{gross} D.H.E. Gross and B.H. Sa, Nucl. Phys. {\bf A437},
	643 (1985); H.R. Jaqaman and D.H.E. Gross,
	Nucl. Phys. {\bf A524}, 321 (1991).

\bibitem{ayik} S. Ayik and C. Gregoire,
	Nucl. Phys. {\bf A513}, 187 (1990).

\bibitem{chase} K. C. Chase and A. Z. Mekjian,
	Phys. Rev. {\bf C49}, 2164 (1994).

\bibitem{finn} J. E. Finn,  et al., Phys. Rev. Lett {\bf 49}, 1321 (1982)..

\bibitem{li1} T. Li, et al., Phys. Rev.  {\bf C49}, 1630 (1994).

\bibitem{li2} T. Li, et al., Phys. Rev. Lett, {bf 70}, 1924 (1993)

\bibitem{coniglio} A. Coniglio and W. Klein, J. Phys. A: {\bf 13}, 2775 (1980).

\bibitem{fisher} M.E. Fisher, Physics, {\bf 3}, 255 (1967).

\bibitem{stinchcombe} R. B. Stinchcombe, in: Phase transitions and
 	critical phenomena, vol. 7, eds. C. Domb and J. L. Lebowitz
	(academic press, London, 1983)

\bibitem{jacak} B.V. Jacak et.al., Phys. Rev. {\bf C35}, 1751 (1987)

\bibitem{ogilvie} C.A. Ogilvie et. al., Phys. Rev. Lett. {\bf 67}, 1214 (1991)

\bibitem{dasgupta} S. Das Gupta and A. Z. Mekjian, Phys. Rept. {\bf 72}, 131
(1981)
\end{thebibliography}
\end{document}